\documentclass[superscriptaddress,twocolumn,10pt,prl,aps,showpacs,longbibliography,floatfix]{revtex4-2}

\usepackage[dvipsnames]{xcolor}

\definecolor{darkred}{rgb}{0.6,0.05,0.05}
\definecolor{darkgreen}{rgb}{0.05,0.6,0.05}
\definecolor{darkblue}{rgb}{0.05,0.05,0.6}
\usepackage[colorlinks]{hyperref}
\hypersetup{colorlinks,
linkcolor=darkred,
filecolor=darkgreen,
urlcolor=darkblue,
citecolor=darkgreen}

\usepackage{titlesec}
\titleformat{\section}[runin]
{\normalfont\itshape}{\thesection}{0em}{}[.]

\usepackage{mathtools}
\usepackage{amsmath}
\allowdisplaybreaks
\usepackage{ bbold }
\usepackage{amssymb}
\usepackage[normalem]{ulem}
\usepackage{amsthm}
\usepackage{xfrac}
\usepackage{dsfont}
\usepackage{siunitx}
\usepackage{physics}
\usepackage{graphicx}% Include figure files
\usepackage{hyperref}% add hypertext capabilities
\usepackage[capitalise]{cleveref}
\usepackage{dcolumn}% Align table columns on decimal point
\usepackage{bm}% bold math
\usepackage{layouts}
\usepackage{comment}

\begin{document}

\author{Filippo Ferrari}
\email{filippo.ferrari@epfl.ch}
\affiliation{Laboratory of Theoretical Physics of Nanosystems (LTPN), Institute of Physics, \'{E}cole Polytechnique F\'{e}d\'{e}rale de Lausanne (EPFL), 1015 Lausanne, Switzerland}
\affiliation{Center for Quantum Science and Engineering, \\ \'{E}cole Polytechnique F\'{e}d\'{e}rale de Lausanne (EPFL), CH-1015 Lausanne, Switzerland}
\author{Vincenzo Savona}
\email{vincenzo.savona@epfl.ch}
\affiliation{Laboratory of Theoretical Physics of Nanosystems (LTPN), Institute of Physics, \'{E}cole Polytechnique F\'{e}d\'{e}rale de Lausanne (EPFL), 1015 Lausanne, Switzerland}
\affiliation{Center for Quantum Science and Engineering, \\ \'{E}cole Polytechnique F\'{e}d\'{e}rale de Lausanne (EPFL), CH-1015 Lausanne, Switzerland}
\author{Fabrizio Minganti}
\email{fabrizio.minganti@gmail.com}
\affiliation{Laboratory of Theoretical Physics of Nanosystems (LTPN), Institute of Physics, \'{E}cole Polytechnique F\'{e}d\'{e}rale de Lausanne (EPFL), 1015 Lausanne, Switzerland}
\affiliation{Center for Quantum Science and Engineering, \\ \'{E}cole Polytechnique F\'{e}d\'{e}rale de Lausanne (EPFL), CH-1015 Lausanne, Switzerland}
\affiliation{Alice \& Bob, 53 boulevard du G\'en\'eral Martial Valin, 75015 Paris, France}

\title{Chaos and thermalization in open quantum systems}

\date{\today}

\begin{abstract}
The eigenstate thermalization hypothesis (ETH) provides a cornerstone for understanding thermalization in isolated quantum systems, linking quantum chaos with statistical mechanics. In this work, we extend the ETH framework to open quantum systems governed by Lindblad dynamics. We introduce the concept of Liouvillian stripe—spectral subset of the non-Hermitian Liouvillian superoperator—which enables the definition of effective pseudo-Hermitian Hamiltonians. This construction allows us to conjecture a Liouvillian version of ETH, whereby local superoperators exhibit statistical properties akin to ETH in closed systems. We substantiate our hypothesis using both random Liouvillians and a driven-dissipative quantum spin chain, showing that thermalization manifests through the suppression of coherent oscillations and the emergence of structureless local dynamics. These findings have practical implications for the control and measurement of many-body open quantum systems, highlighting how chaotic dissipative dynamics can obscure the system response to external probes. 
\end{abstract}

\maketitle

%%%%%%%%%%%%%%%%%%%%%%%%%%%%%%%%%%%%%%%%%%%%%%%%%%%%%%%%%

\section{Introduction}\label{sec:introduction}
Statistical mechanics is a pillar of modern physics. It describes the macroscopic properties of a many-body system in contact with a thermal bath using a few physical quantities, such as temperature or chemical potential.
The predictions of statistical mechanics often extend to isolated systems that, after a transient, behave as thermodynamic ones.
In classical mechanics, thermalization occurs by ergodicity: any available configuration in the phase space is eventually explored if one waits for a long enough time.
Although strictly speaking not related, in most systems, ergodicity is accompanied by chaos \cite{strogatz_nonlinear_2018}. 
A chaotic system is one where ``an approximate knowledge of the present does not guarantee an approximate knowledge of the future,'' as any perturbation on an initial condition is exponentially amplified along the dynamics \cite{lorenz_deterministic_1963}.
Chaos and thermalization occur in classical \textit{nonlinear} systems.

Quantum systems can thermalize, even if quantum mechanics is a linear theory. 
A thermal density matrix describes their averaged, local properties despite the whole system remaining in a pure state.
An explanation of this phenomenon comes from the eigenstate thermalization hypothesis (ETH) \cite{deutsch_quantum_1991, srednicki_chaos_1994, srednicki_approach_1999, rigol_thermalization_2008, dalessio_quantum_2016,Deutsch2018}: individual energy eigenstates average to the statistical ensemble with the same average energy.
The ETH can be regarded, very broadly, as the quantum manifestation of ergodicity that, together with chaos, describes how statistical mechanics emerges.
Quantum chaos manifests itself through the spectral properties of the Hamiltonian, whose features conform to the predictions of random matrix theory (RMT) \cite{guhr_random-matrix_1998, haake_quantum_2010, dalessio_quantum_2016}.

Not all quantum systems are isolated, and the exchanges between the system and its environment may not lead to thermal equilibrium. 
These open quantum systems do not obey a Hamiltonian evolution, but require the introduction of more general quantum maps \cite{breuer_theory_2007}.
In analogy with closed quantum systems, chaos can be defined through spectral properties of the generator of the dynamics.
In particular, the RMT predictions have been extended to systems described by the Lindblad master equation, describing quantum systems interacting with a Markovian bath, highlighting similarities and differences with respect to its classical counterpart~\cite{akemann_universal_2019, hamazaki_universality_2020, sa_complex_2020, dahan_classical_2022, prasad_dissipative_2022, kawabata_symmetry_2023, villasenor_breakdown_2024, ferrari_dissipative_2025}.
These systems unavoidably evolve towards their (often unique) steady state. 
Concepts such as ergodicity, thermalization, or chaos are often described as transient phenomena or 
as properties of the steady state.
This interpretation is in contrst with the description of chaos and thermalization as persisting in the long-time dynamics of either classical or quantum systems after, e.g., the occurrence of prethermalization phenomena \cite{bertini_prethermalization_2015, mori_thermalization_2018}.
Whence the question: can we define an analogue of the ETH hypothesis for open quantum systems? And what is its meaning?

Here, we extend the ETH hypothesis to open quantum systems described by a Lindblad master equation.
Key to this understanding is the dynamics of superoperators and the analogy between portions of the Liouvillian spectrum and the spectrum of a non-Hermitian Hamiltonian.
Taking advantage of the peculiar random matrix structure emerging in chaotic systems, and applying results in non-Hermitian physics, we tie our hypothesis to the ETH characterizing Hamiltonian quantum systems.
We corroborate our conjecture through two examples.
Furthermore, in systems characterized by Liouvillian ETH, we show that local observables exhibit suppressed oscillations around a well-defined average value, in analogy with thermalization in Hamiltonian systems.
In turn, this implies that signals acquired by probing the system are structureless, possibly hindering the control and operation of many-body open quantum systems.

\section{Thermalization in quantum systems}\label{sec:open_ETH}

Consider an extended system described by a Hamiltonian $\hat{H}$ with eigenstates $\ket{\Psi_\alpha}$.
Let $\hat{O}$ be a local operator; i.e.,  $\hat{O}$ acts at a specific location in space.
$O_{\alpha\beta} = \bra{\Psi_\alpha}\hat{O}\ket{\Psi_\beta}$ are the matrix elements of
$\hat{O}$.
The eigenstate thermalization hypothesis states that \cite{Deutsch2018}
\begin{equation}\label{eqs:ETH}
    O_{\alpha\beta} = \widetilde{O}(\overline{E})\delta_{\alpha\beta} + e^{-S(\overline{E})/2}f(\overline{E}, \omega)R_{\alpha\beta},
\end{equation}
where $\widetilde{O}(\overline{E})$ and $f(\overline{E}, \omega)$ are smooth functions of their arguments, $2\overline{E} = E_\alpha + E_\beta$ and $\omega=E_\alpha-E_\beta$, $S = \kappa_b \log(\Omega)$ is the thermodynamic entropy associated with number of explorable states, and $R_{\alpha\beta}$ are normally-distributed random numbers.
For a random matrix $\hat{H}$, the ETH predicts
\begin{equation}\label{eqs:ETH_rm}
    O_{\alpha\beta} = \overline{O}\,\delta_{\alpha\beta} + \sqrt{\frac{\overline{O^2}}{D}}\,R_{\alpha\beta},
\end{equation}
where $\overline{O}=\sum_a\bra{a}\hat{O}\ket{a}/D$ and $D$ is the dimension of the Hamiltonian.

The ETH Ansatz has been extended to non-Hermitian Hamiltonians \cite{singha_roy_unveiling_2025}, an effective formalism with various applications in optical systems with gain and losses, cold gases, and weakly-nonlinear quantum systems \cite{Moiseyev2011,Bender2007}.
Non-Hermitian Hamiltonians admit left and right eigenvectors. 
For non-Hermitian random matrices, ETH holds only for the right eigenvectors $\ket{\phi_\alpha}$, reading \cite{cipolloni_non-hermitian_2024, singha_roy_unveiling_2025}
\begin{equation}\label{eqs:ETH_nhrm}
    O_{\alpha\beta} = \overline{O}\,\langle\phi_\alpha|\phi_\beta\rangle + \sqrt{\frac{\overline{O^2}}{D}}\,R_{\alpha\beta}.
\end{equation}

That of non-Hermitian Hamiltonians is, however, a formalism that cannot be extended to all open quantum systems. In the Markov approximation, these are governed by the Lindblad master equation \cite{breuer_theory_2007} for the reduced density matrix $\hat{\rho}$
\begin{equation}\label{eqs:lindblad}
    \frac{\partial \hat{\rho}}{\partial t} = \mathcal{L}\hat{\rho} = -i[\hat{H}, \hat{\rho}] + \sum\gamma_j\left(\hat{L}_j\hat{\rho}\hat{L}_j^\dagger -\frac{1}{2}\{\hat{L}_j^\dagger\hat{L}_j,\hat{\rho}\}\right),
\end{equation}
where $\hat{L}_j$ are jump operators (with dissipation rates $\gamma_j$) describing the interaction between system and its environment.
$\mathcal{L}$ is the Liouvillian, the non-Hermitian superoperator that generates the dynamics.
The Liouvillian can be diagonalized as
\begin{equation}
    \mathcal{L}\hat{\eta}_\alpha = \lambda_\alpha \hat{\eta}_\alpha, \quad \lambda_\alpha = \Gamma_\alpha + i\Omega_\alpha,
\end{equation}
with $\hat{\eta}_\alpha$ the right eigenstates and $\lambda_\alpha$ the eigenvalues.
Aside from exceptional points, the set $\hat{\eta}_\alpha$ forms a basis of the space of density matrices.

We consider Liouvillians whose spectral structure follows the predictions of non-Hermitian random matrix ensembles \cite{grobe_quantum_1988}: Ginibre distribution
for chaotic systems; Poisson 2D for integrable ones.

\section{Liouvillian stripes and associated Hamiltonian}
We introduce the concept of a vertical stripe of the spectrum of a non-Hermitian matrix. 
Each stripe is defined by fixing the real part $\overline{\Gamma}$, so that each eigenvalue can be expressed as $\lambda_\alpha = \overline{\Gamma} \pm d_{\rm max}/2 + i\Omega_\alpha$, with $d_{\rm max}\ll \textrm{max}(|\Omega_\alpha|)$. 
Details on the choice of $d_{\rm max}$ can be found in the Supplementary Material.
We then analyze the spectral properties of each stripe.
For a 2D Possion distribution of eigenvalues (an integrable non-Hermitian system) the stripe displays the 1D Poissonian distribution associated with integrable Hermitian systems.
For a Ginibre distribution (a chaotic system) the stripe shows the Wigner-Dyson-like distribution of Hermitian random matrices.

Considering Liouvillian stripes, these contain eigenvalues that, upon removal of the decaying part $\bar{\Gamma}$, are purely imaginary and symmetrically distributed around zero.
Therefore, the dynamics within a stripe is captured by a reduced operator $\hat{H}_{\rm eff}(\overline{\Gamma})$ that is pseudo-Hermitian \cite{MOSTAFAZADEH2010}. 
This effective Hamiltonian is a matrix of size $n_s \times n_s$ with purely real eigenenergies $\Omega_\alpha$.
These findings, supported by the data presented in the Supplementary Information, lead us to formulate the following conjecture.

\section{ETH conjecture in open quantum systems and physical interpretation}

Assume $\hat{H}_{\rm eff}$ describes a physical model.
It now makes sense to study ETH by investigating how local operators behave on the eigenstates of $\hat{H}_{\rm eff}$.
However, local operators on $\hat{H}_{\rm eff}(\overline{\Gamma})$ are superoperators when described in the full Liouvillian space.
We therefore conclude that there is an analogy between the occurrence of ETH for \textit{superoperators} acting on a \textit{Liouvillian stripe} and ETH in non-Hermitian systems.
We thus conjecture that the eigenstate thermalization hypothesis holds for local superoperators, thus generalizing the ETH Ansatz to open quantum systems, upon the introduction of a Liouvillian stripe.
Within a sufficiently small energy window $\omega = \Omega_\alpha - \Omega_\beta$, an average energy $ 2 \overline{E} = \Omega_\alpha + \Omega_\beta$, and around a given decay rate $\overline{\Gamma}$, fixed by the Liouvillian stripe, we have
\begin{equation}\label{eqs:SUPER}
    \mathcal{O}_{\alpha\beta} = \overline{\mathcal{O}}(\overline{E}, \overline{\Gamma})\delta_{\alpha\beta} + e^{-S(\overline{E})/2}f(\overline{E}, \omega, \overline{\Gamma})R_{\alpha\beta}.
\end{equation}
Compared to a straightforward generalization of ETH to Lindbladians, the fact that each stripe defines a different $\hat{H}_{\rm eff}(\overline{\Gamma})$ manifests in the dependence of the microcanonical prediction $\overline{\mathcal{O}}(\overline{E}, \overline{\Gamma})$ as well as on the smooth function $f(\overline{E}, \omega, \overline{\Gamma})$.

In Hamiltonian systems, thermalization entails the fact that the expectation values of local observables on each eigenstate averages to the same value of the corresponding statistical ensemble.
It follows that thermalization on a stripe must lead to a similar emergent behavior, where the local properties must average out through the various frequencies $\Omega_\alpha$.
This loss of coherence and the lack of oscillation in the decay towards the steady state is not captured by $\overline{\Gamma}$, the real part of the Liouvillian eigenvalues.

Although any object that acts on the space of density matrices is a superoperator, like $\mathcal{L}$, 
we focus here on two specific kinds: coherent and measurement superoperators defined by \footnote{For numerical purposes, these can be defined as $\mathcal{O}^{\rm coh} = \hat{O}\otimes\mathds{1} - \mathds{1}\otimes\hat{O}^{\mathsf{T}}$ and $\mathcal{O}^{\rm m} = \hat{O}\otimes\hat{O}^*$.}
\begin{equation}\label{eqs:superoperators}
    \begin{split}
        &\mathcal{O}^{\rm coh} \hat{\rho} = [\hat{O},\,\hat{\rho}\,], \qquad \mathcal{O}^{\rm m} \hat{\rho}= \hat{O} \, \hat{\rho} \,\hat{O}^\dagger.
    \end{split}
\end{equation}
$\mathcal{O}^{\rm coh}$ represents a coherent perturbation of an open quantum system via an operator $\hat{O}$. It captures the linear response of a system in, e.g., a pump-and-probe experiment.
$\mathcal{O}^{\rm m}$ represents the result of a generalized measurement, with $\hat{O}$ acting as an element of a Kraus map.
Thus, these superoperators can be connected with coherent and incoherent probing of the system.

\section{Example I: random Liouvillian}\label{sec:results}

\begin{figure}[t!]
\centering
\includegraphics[width=0.45\textwidth]{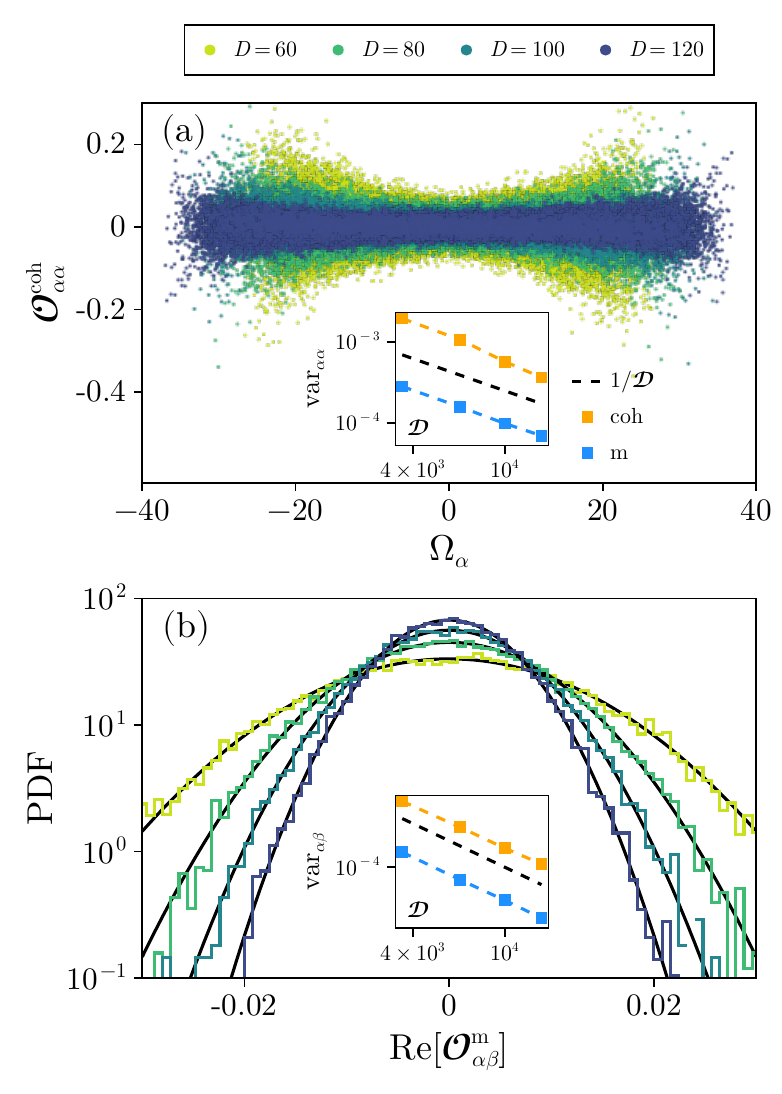}
\caption{Signatures of ETH in random Liouvillians. 
(a) As a function of the stripe energies $\Omega_\alpha$, diagonal matrix elements of the coherent superoperator $\mathcal{O}^{\rm coh}_{\alpha\alpha}$ in Eq.~\eqref{eqs:superoperators} with $\hat{O} = \hat{\sigma}^x\otimes\mathds{1}_{\mathcal{D}/2}$, for different Hilbert space dimensions and various Liouvillian stripes, as detailed below.
The inset shows the scaling of the variance $\textrm{var}_{\alpha\alpha}$ of both the coherent (orange line) and measurement (blue line) superoperators, compared to $1/\mathcal{D}$ with $\mathcal{D}$ the size of the Liouvillian.
The variance is computed over the diagonal matrix elements such that $|\Omega_\alpha|<10$.
(b) Distribution of the off-diagonal matrix elements of the measurement superoperator $\mathcal{O}^{\rm m}_{\alpha\beta}$ in Eq.~\eqref{eqs:superoperators}, with the same $\hat{O}$ and system sizes analyzed in panel (a), and the procedure detailed in the main text.
% The off-diagonal elements are selected on $N_s$ Liouvillian stripes and $N_{\rm dis}$ realizations of the disorder for energy differences within the window $\omega\pm\delta\omega$ with $\omega=10$ and $\delta\omega=0.2$. 
Each probability density function (PDF) for $\mathcal{O}^{\rm m}_{\alpha\beta}$ and a given $\mathcal{D}$ is compared with a Gaussian distribution (black lines) with variance $\textrm{var}_{\alpha\beta}$.
The inset shows the scaling of $\textrm{var}_{\alpha\beta}$ of both the coherent (orange line) and measurement (blue line) superoperators, compared to $1/\mathcal{D}$.
The random Liouvillian parameters are fixed to $r=6$, $\beta=2$, $g_{\rm eff} = 1.05$ \cite{Note2}.
To keep the total number of eigenstates of order $10^5$, we choose:
$D=60$ ($N_{\rm dis}=56$ realizations, $N_s=12$ stripes); $D=80$ ($N_{\rm dis}=32$ realizations, $N_s=14$ stripes); $D=100$ ($N_{\rm dis}=20$ realizations, $N_s=16$ stripes); and $D=120$ ($N_{\rm dis}=14$ realizations, $N_s=18$ stripes). 
}
\label{fig:ETH_rl}
\end{figure}

We construct a random Liouvillian $\mathcal{L}$~\cite{denisov_universal_2019, sa_spectral_2020}, depending on three parameters $r$, $\beta$, and $g_{\rm eff}$, on a Hilbert space of dimension $D$ \footnote{Following Ref.~\cite{sa_spectral_2020}
, we define a basis for the operators in the Hilbert space $\{\hat G_j\}$ with $j=0, \ldots, D^2-1$ such that $\operatorname{Tr}[\hat G_j^{\dagger} \hat G_k]=\delta_{j k}$, with $\hat G_0= \mathbb{1} / \sqrt{D}$. We construct $r$ jump operators as $\hat L_{j}= g \sum_{k=1}^{D^2-1} \hat G_k w_{k, j}$, with $\hat L_{j}$ traceless and $w$ a matrix sampled from a Ginibre ensemble.
The Hamiltonian $\hat{H}$ is a random matrix sampled from the Gaussian unitary ensemble. We then define $\mathcal{L}$ according to Eq.~\eqref{eqs:lindblad}. The coupling parameter is given in terms of the rescaled parameter $g_{\rm eff} = (2r\beta N)^{1/4}g$}.
We collect a series of Liouvillian stripes within the spectral bulk (see the Supplmentary Information).
Over each Liouvillian stripe, and over many random realizations, we study the matrix elements of the superoperators in Eq.~\eqref{eqs:superoperators} with $\hat{O} =  \hat{\sigma}^x\otimes\mathds{1}_{D/2}$, being $\mathds{1}_{D/2}$ the $D/2\times D/2$ identity operator.
In Fig.~\ref{fig:ETH_rl} (a) we plot the diagonal matrix elements of the coherent superoperator $\mathcal{O}^{\rm coh}_{\alpha\alpha}$ over several Liouvillian stripes, as a function of the energies $\Omega_\alpha$, for random Liouvillians with increasing Hilbert space dimensions.
The spread of $\mathcal{O}^{\rm coh}_{\alpha\alpha}$ decreases as the inverse of the Liouville space dimension $\mathcal{D} = D^2$, with $D$ the dimension of the Hilbert space.
The inset of Fig.~\ref{fig:ETH_rl} (a) shows, as a function of the system size, the scaling of the variance $\textrm{var}_{\alpha\alpha}$ of the diagonal elements of both the coherent and measurement superoperators restricted to the $\mathcal{O}_{\alpha\alpha}$ such that $|\Omega_\alpha|<10$. 
Data are compared with theoretical prediction $\sim 1/\mathcal{D}$.

Next, we analyze the off-diagonal matrix elements of the measurement superoperator $\mathcal{O}_{\alpha\beta}^{\rm m}$.
We focus on the statistical behavior of $\mathcal{O}_{\alpha\beta}^{\rm m}$ for energies such that $|\omega|=|\Omega_\alpha-\Omega_\beta|=10 \pm \delta\omega$, within $\delta\omega = 0.2$. 
Similar results are obtained for different choices of $\omega$ and $\delta\omega$ (not shown).
In Fig.~\ref{fig:ETH_rl} (b) we show that the statistical distribution of $\mathcal{O}_{\alpha\beta}^{\rm m}$ approximates a Gaussian with zero mean and variance $\textrm{var}_{\alpha\beta}$.
The inset of Fig.~\ref{fig:ETH_rl} (b) shows the scaling of $\textrm{var}_{\alpha\beta}$ as a function of $\mathcal{D}$, for both the coherent and measurement superoperators.
Data are again compared with the theoretical prediction $\sim 1/\mathcal{D}$.

Overall, the analysis demonstrates the validity of the Liouvillian ETH for random Lindblad dynamics.

\section{Example II: driven-dissipative quantum spin chain}\label{sec:results}

We consider a driven-dissipative quantum spin chain whose Hamiltonian is that of the XXZ model with a single magnetic impurity, reading
\begin{equation}\label{eqs:spin_chain_1_ham}
    \hat{H} = J\sum_{j=1}^{N-1}\left(\hat{\sigma}_j^x\hat{\sigma}_{j+1}^x + \hat{\sigma}_j^y\hat{\sigma}_{j+1}^y + \Delta\hat{\sigma}_j^z\hat{\sigma}_{j+1}^z\right) + h\hat{\sigma}_{N/2}^z.
\end{equation}
The system is subject to boundary incoherent drive and dissipation, and dephasing on each site, reading
\begin{align}\label{eqs:spin_chain_1_diss}
    &\hat{L}_1^+ = \sqrt{\gamma^+_1}\hat{\sigma}_1^+,\qquad\hat{L}_1^- = \sqrt{\gamma^-_1}\hat{\sigma}_1^-,\\
    \hat{L}_N^+ = \sqrt{\gamma^+_N}&\hat{\sigma}_N^+,\qquad\hat{L}_N^- = \sqrt{\gamma^-_N}\hat{\sigma}_N^-,\qquad\hat{L}_j^z = \sqrt{\gamma^z}\hat{\sigma}^z_j\nonumber.
\end{align}
The Hamiltonian model alone at $h=0$, i.e., without impurity, is integrable.
For $h\neq 0$, instead, it is chaotic and satisfies Hermitian ETH \cite{brenes_eigenstate_2020, brenes_low-frequency_2020}.
For $\Delta=0$ and $h=0$, the full Liouvillian can be mapped onto the 1D Fermi-Hubbard model with an imaginary interaction that can also be exactly solved with the Bethe ansatz \cite{medvedyeva_exact_2016}.
At nonzero $\Delta$ and $h$, instead, the Liouvillian is quantum chaotic.
The Liouvillian described by Eqs.~\eqref{eqs:spin_chain_1_ham} and \eqref{eqs:spin_chain_1_diss} commutes with the total magnetization superoperator $\mathcal{S}^z = \hat{S}^z\otimes\mathds{1} - \mathds{1}\otimes\hat{S}^z$, being $\hat{S}^z=\sum_j\hat{\sigma}^z_j$ the magnetization operator.
This weak $\mathbb{U}(1)$ Liouvillian symmetry allows the block diagonalization of $\mathcal{L}$.
Here, we focus on the zero super-magnetization sector, for which all eigenstates obey $\mathcal{S}^z \hat{\eta}_j =0$.

\begin{figure}[t!]
\centering
\includegraphics[width=0.5\textwidth]{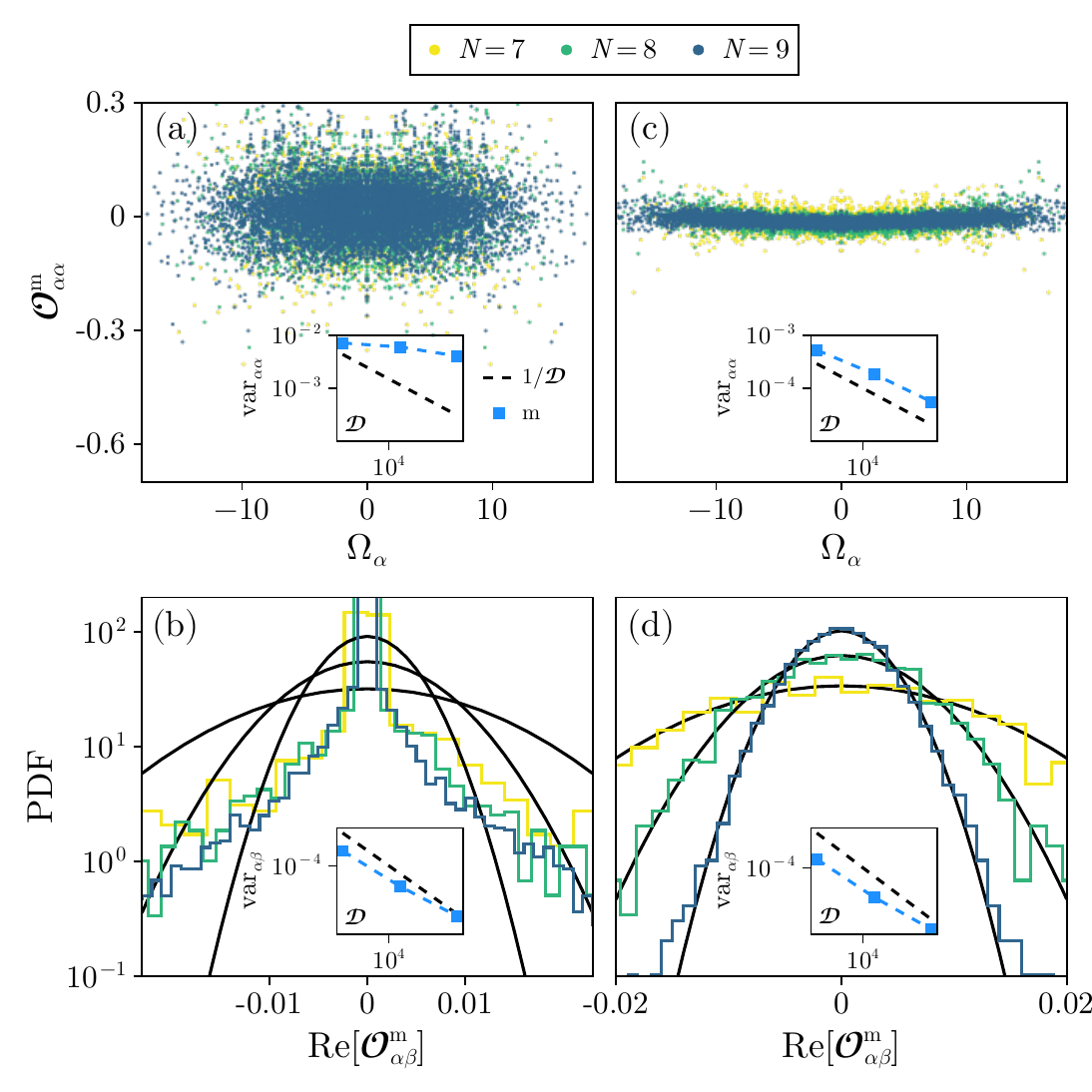}
\caption{Signatures of Liouvillian ETH in a driven-dissipative quantum spin chain. 
We investigate the measurement superoperator $\mathcal{O}^{\rm m}_{\alpha\alpha}$, with $\hat{O} = \hat{\sigma}^x_3$, defined in Eq.~\eqref{eqs:superoperators} and system sizes $N=7, 8, 9$ (from light green to dark blue). 
(a) and (b) Integrable Liouvillian with $\Delta=h=0$.
(a) Diagonal matrix elements for $N_s=19$ Liouvillian stripes as a function of the stripe energies $\Omega_\alpha$.
The inset shows the scaling of the variance $\textrm{var}_{\alpha\alpha}$ of the measurement superoperator, compared to $1/\mathcal{D}$ (black-dashed line) as a function of the dimension of the Liouvillian, $\mathcal{D}$.
The variance is computed over the diagonal matrix elements such that $|\Omega_\alpha|<5$.
(b) Distribution of the off-diagonal matrix elements $\mathcal{O}^{\rm m}_{\alpha\beta}$.
The off-diagonal elements are selected on $N_s=19$ Liouvillian stripes for energy differences within the window $|\omega|\pm\delta\omega$ with $\omega=8$ and $\delta\omega=0.1$. 
Each PDF for $\mathcal{O}^{\rm m}_{\alpha\beta}$ and a given $N$ is compared with a Gaussian distribution (black lines) with variance $\textrm{var}_{\alpha\beta}$.
The inset shows the scaling of $\textrm{var}_{\alpha\beta}$ of the measurement superoperator, compared to $1/\mathcal{D}$ as a function of $\mathcal{D}$.
(c) and (d) Same as in panels (a) and (b) but for the chaotic Liouvillian with $\Delta=0.8$ and $h=1$.
Other parameters are fixed to $J=1$, $\gamma^z=1$, $\gamma_1^+ = 0.5$, $\gamma_1^-=1.2$, $\gamma_N^+=1$, $\gamma_N^-=0.8$.
}
\label{fig:ETH_spin1}
\end{figure}

In the following discussion, we will consider the local operator $\hat{O} = \hat{\sigma}^x_3$ (qualitatively similar results are obtained for other choices of $\hat{O}$) and focus on the measurement superoperator $\mathcal{O}^{\textrm{m}} = \hat{\sigma}_3^x\otimes\hat{\sigma}_3^x$. 
We construct a series of Liouvillian stripes in the spectral bulk (see the Supplementary Information) and, to gather statistics, results are averaged over $N_s$ contiguous  Liouvillian stripes, where we verified that the functions $\overline{\mathcal{O}}(\overline{E}, \overline{\Gamma})$ and $f(\overline{E}, \omega, \overline{\Gamma})$ defined in Eq.~\eqref{eqs:SUPER} are almost constant.

First, we focus on the itegrable case $\Delta=h=0$.
In Fig.~\ref{fig:ETH_spin1} (a) we plot the diagonal matrix elements of $\mathcal{O}^{\rm m}_{\alpha\alpha}$ as a function of the energies $\Omega_\alpha$.
The variance $\textrm{var}_{\alpha\alpha}$ of $\mathcal{O}^{\rm m}_{\alpha\alpha}$ is almost constant with the system size (see the inset), indicating the failure of the thermalization hypothesis.
Similarly, the distribution of the off-diagonal matrix elements $\mathcal{O}_{\alpha\beta}^{\rm m}$ shows significant deviations from a Gaussian profile for all the system sizes we analyzed, as shown in Fig.~\ref{fig:ETH_spin1} (b).
We thus confirm that an integrable model does not follow the prediction of Liouvillian ETH.

Then, we study the chaotic case with $\Delta=0.8$ and $h=1$. 
The diagonal matrix elements $\mathcal{O}^{\rm m}_{\alpha\alpha}$ are shown in Fig.~\ref{fig:ETH_spin1} (c), and they shrink with the system size. The variance $\textrm{var}_{\alpha\alpha}$ scales as $\sim1/\mathcal{D}$ (see the inset), being $\mathcal{D}=\binom{2N}{N}$ the dimension of the Liouvillian block of zero magnetization, in agreement with the ETH Ansatz.
The results presented in Fig.~\ref{fig:ETH_spin1} (d) confirm that the off-diagonal elements $\mathcal{O}_{\alpha\beta}^{\rm m}$ are normally distributed with variance $\textrm{var}_{\alpha\beta}$, for $|\omega|=8$ and $\delta\omega=0.1$.
Similar results are obtained for different choices of $\omega$ and $\delta\omega$ (not shown).
The inset of Fig.~\ref{fig:ETH_spin1} (d) shows that the scaling of $\textrm{var}_{\alpha\beta}$ follows $\sim 1/\mathcal{D}$.
Furthermore, we test that, considering different observables and different stripes, leads to different distributions and values of the variances of both $\mathcal{O}^{\rm m}_{\alpha\alpha}$ and $\mathcal{O}^{\rm m}_{\alpha\beta}$.
We conclude that the functions $\overline{\mathcal{O}}(\overline{E}, \overline{\Gamma})$ and $f(\overline{E}, \omega, \overline{\Gamma})$ are indeed dependent on the selected stripe.

\begin{figure}[t!]
\centering
\includegraphics[width=0.45\textwidth]{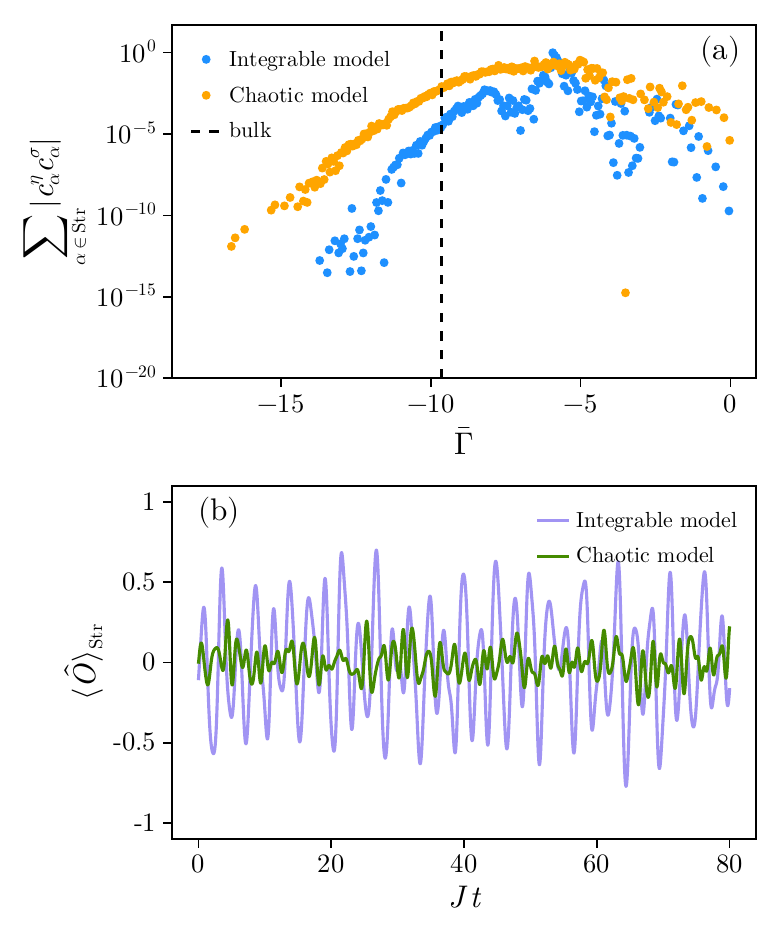}
\caption{Liouvillian ETH and dynamics. (a) Sum of the spectral coefficients over single stripes, $\sum_{\alpha\in\textrm{Str}}|c^\eta_\alpha c^\sigma_\alpha|$, for the integrable (blue dots) and chaotic (orange dots) model. 
Each dot represents a single stripe and the width of the stripe is chosen according to the procedure described in the Supplementary Information.
(b) Renormalized dynamics of the operator $\hat{O}=i(\hat{\sigma}^+_1\hat{\sigma}^-_N - \hat{\sigma}^+_N\hat{\sigma}^-_1)$ over a stripe, according to Eq.~\eqref{eqs:dynaics_stripe} at a $\overline{\Gamma}$ marked by the black-dashed line in panel (a). 
The purple line referes to the integrable model, while the green line indicates the chaotic model.
The values of $\Delta$ and $h$ are both 0 for the integrable model and $\Delta=0.8$ and $h=1$ for the chaotic model.
The other parameters are fixed as in Fig.~\ref{fig:ETH_spin1}.
}
\label{fig:dynamics}
\end{figure}

Finally, we study the dynamical consequences of Liouvillian ETH in the spin chain in Eqs.~\eqref{eqs:spin_chain_1_ham} and \eqref{eqs:spin_chain_1_diss} with $N=8$ spins.
The expectation value of an operator $\hat{O}$, $\langle\hat{O}\rangle(t) = \operatorname{Tr}[\hat{O}\hat{\rho}(t)]$ can be expanded as
\begin{equation}\label{eqs:dynaics}
    \langle\hat{O}\rangle(t) = \sum_\alpha e^{\lambda_\alpha t} c_\alpha^\eta c_\alpha^\sigma
\end{equation}
where $c_\alpha^\eta = \langle \hat{O}^\dagger|\hat{\eta}_\alpha\rangle$
is the projection of $\hat{O}$ over $\hat{\eta}_\alpha$ and $c_\alpha^\sigma = \langle \hat{\sigma}_\alpha|\hat{\rho}(0)\rangle$ are the projection of the initial density matrix $\hat{\rho}(0)$ over the left Liouvillian eigenstates $\hat{\sigma}_j$, with the scalar product defined by $\langle\hat{A}| \hat{B}\rangle = \operatorname{Tr}[\hat{A}^\dagger \hat{B}]$.
The operator we consider is $\hat{O} = i(\hat{\sigma}^+_1\hat{\sigma}^-_N - \hat{\sigma}^+_N\hat{\sigma}^-_1)$, measuring the current fluctuations at the driven edges of the chain (see the Supplementary Information for the dynamics of other operators).
As initial state we choose the density matrix $\hat{\rho}(0) = \ketbra{\uparrow\,...\,\uparrow}$, belonging to the zero-magnetization symmetry sector.
In Fig.~\ref{fig:dynamics} (a) we plot the sum of the spectral weights on the stripe, $\sum_{\alpha\in \textrm{Str}}|c_\alpha^\eta c_\alpha^\sigma|$, as a function of the stripe decay rate $\overline{\Gamma}$.
We observe that $\hat{O}$ has a support over the bulk of the Liouvillian spectrum for both the chaotic and integrable limits of the model, indicated by the smooth and slow decrease of the spectral weights.
In Fig.~\ref{fig:dynamics} (b) we plot the dynamics of the observable over a single stripe deep in the spectral bulk [$\overline{\Gamma}$ marked by the black-dashed line in Fig.~\ref{fig:dynamics} (a)].
To compare the integrable and chaotic model, the expectation value has been renormalized by the total spectral weight,
\begin{equation}\label{eqs:dynaics_stripe}
    \langle\hat{O}\rangle_{\rm Str}(t) = \frac{\sum_{\alpha\in\textrm{Str}}e^{i\Omega_\alpha t}c^\eta_\alpha c_\alpha^\sigma}{\sum_{\alpha\in\textrm{Str}}|c^\eta_\alpha c_\alpha^\sigma|}.
\end{equation}
For the integrable model, we find that only a few spectral weights within the stripe are larger than $10^{-10}$, thus contributing to the dynamics. 
This produces large oscillations in the dynamics of $\langle\hat{O}\rangle_{\rm Str}$ (purple line), which can be regarded as a consequence of the integrability of the spin chain.
For the chaotic model, instead, the spectral weights are relevant over the full stripe.
We notice that an extensive number of eigenvectors involved in the dynamics was also a necessary condition to define steady-state chaos in open quantum systems \cite{ferrari_dissipative_2025}.
The mixing of the frequencies $\Omega_\alpha$, weighted by the projections of the chaotic eigenstates $\ket{\hat{\eta}_\alpha}$ over $\hat{O}$, causes an evident damping in the oscillations of $\langle\hat{O}\rangle_{\rm Str}$ (green line).
We verified that the results hold for the stripes in the spectral bulk.
The Liouvillian ETH has therefore two main consequences on the open quantum dynamics.
First, it implies that despite the presence of oscillating properties predicted by the Liouvillian spectrum, these may not be accessible by probing local operators.
From this, it follows that spectral response functions (i.e., the Fourier transform of signals related to a given observable) will be structureless, making it challenging to determine the eigenfrequencies and time scales of the system.
This lack of features closely resembles thermalizing Hamiltonian systems, and draws an additional parallel between Liouvillian and Hamiltonian ETH.

\section{Discussion and conclusions}\label{sec:conclusion}

In this paper we introduced a framework to probe the eigenstate thermalization hypothesis in general Markovian open quantum systems, such as disordered random Liouvillians or interacting driven-dissipative many-body quantum lattices.
By gauging out the negative decay rate of the Liouvillian eigenvalues, we drew the analogy between stripes of $\mathcal{L}$ and chaotic pseudo-Hermitian Hamiltonians for which we assessed the ETH for coherent and measurement superoperators.
We discussed the connection between Lindbladian ETH and the dynamics of observables in integrable and chaotic dissipative spin chains, detailing the physical consequences of thermalization in open quantum systems.

The Hermitian ETH has been recently generalized beyond first and second moments (average and variance) within the language of free probability \cite{pappalardi_eigenstate_2022, fritzsch_microcanonical_2025, fava_designs_2025, pappalardi_full_2025}.
An interesting perspective is the generalization of these results to open Markovian systems.
Other open questions are the study many-body effects  violating ETH, such as many-body localization \cite{oganesyan_localization_2007, pal_many-body_2010, bardarson_unbounded_2012, serbyn_local_2013, abanin_colloquium_2019}, quantum scars \cite{turner_weak_2018, turner_quantum_2018, serbyn_quantum_2021}, and Hilbert space fragmentation \cite{sala_ergodicity_2020, khemani_localization_2020, moudgalya_quantum_2022, brighi_hilbert_2023}, in open system configurations \cite{hamazaki_ARXIV_2022, li_hilbert_2023, garciagarcia_ARXIV_2025}.

\section{Note added}\label{sec:note}
At the final stage of preparation of the manuscript, Ref.~\cite{almeida_ARXIV_2025} was uploaded on the arXiv. This article focuses on an alternative definition of eigenstate thermalization in open quantum systems, not based on superoperators or the definition of a Liouvillian stripe.

\section{Acknowledgements}\label{sec:acknowledgements}

We acknowledge useful discussions with Ievgen I. Arkhipov and Joachim Cohen. This research was funded in whole or in part by the Swiss National Science Foundation (SNSF) through Projects No. 200021-227992, 200020\_215172, and 20QU-1\_215928, and was supported as a part of NCCR SPIN, a National Centre of Competence in Research, funded by the Swiss National Science Foundation (grant number 51NF40-225153).

\clearpage
\onecolumngrid

\begin{center}
  \textbf{\large Supplementary Information for \\ ``Chaos and thermalization in open quantum systems''}\\[1em]
  Filippo Ferrari, Vincenzo Savona and Fabrizio Minganti
\end{center}
\vspace{1em}

\titleformat{\section}{\bfseries\centering\uppercase}{\thesection.}{1em}{}
\titleformat{\subsection}{\bfseries}{\thesubsection}{1em}{}

\renewcommand{\thesection}{\Roman{section}}
\renewcommand{\thesubsection}{\Alph{subsection}}

\section{I. Analysis of Hamiltonian and dissipative quantum chaos}

In quantum mechanics, integrability and chaos can be assessed via the spectral analysis of the real eigenenergies in the Hermitian case, or the complex eigenvalues in the non-Hermitian case.
This is at the heart of the nowadays universally accepted Bohigas-Giannoni-Schmitt \cite{bohigas_characterization_1984} and Grobe-Haake-Sommer \cite{grobe_quantum_1988} conjectures, respectively.
The object of interest is the statistical distribution of the eigenvalue distances, $s_\alpha = E_{\alpha+1}-E_\alpha$ (being $E_\alpha$ the eigenenergies of the Hermitian time evolution generator) or $s_\alpha = |\lambda_\alpha-\lambda_\alpha^{\rm NN}|$ (being $\lambda_\alpha$ the eigenvalues of the non-Hermitian time evolution generator, and $\lambda_\alpha^{\rm NN}$ its nearest neighbor in the complex plane).
In integrable systems, the level statistics follows the Poissonian distribution in one or two dimensions typical of uncorrelated random variables.
In chaotic systems, instead, the statistical distribution of the spacings conforms to Hermitian or non-Hermitian random matrix theory (RMT).
Unveiling the Poissonian or RMT structure of the spectrum requires a procedure know as unfolding \cite{guhr_random-matrix_1998}, where the system dependent part of the spectrum is separated from the universal fluctuations.

Signatures of RMT structure in the spectrum which can be obtained avoiding the unfolding procedure have been identified in both Hermitian and non-Hermitian systems.
For Hermitian systems, one looks at the Hamiltonian ratio \cite{atas_distribution_2013}, defined as
\begin{equation}\label{eqs:ham_ratio}
    r_\alpha = \frac{\textrm{min}\left(s_{\alpha}, s_{\alpha-1}\right)}{\textrm{max}\left(s_{\alpha}, s_{\alpha-1}\right)},
\end{equation}
and one considers the average $\langle r \rangle$ over the $r_\alpha$. 
In the case of Poissonian statistics, one finds $\langle r \rangle\simeq 0.386$, whereas RMT statistics for the Gaussian Orthogonal Ensemble produces $\langle r \rangle\simeq 0.53$.
For non-Hermitian system, the above ratio has been generalized to the complex spacing ratio \cite{sa_complex_2020}
\begin{equation}\label{eqs:complex_ratio}
    z_\alpha = \frac{\lambda_\alpha - \lambda_\alpha^{\rm NN}}{\lambda_\alpha - \lambda_\alpha^{\rm NNN}},
\end{equation}
where $\lambda_\alpha^{\rm NNN}$ is the next-to-nearest neighbor of $\lambda_\alpha$.
The distribution of the $z_\alpha$ in the complex plane distinguishes Poissonian from non-Hermitian RMT structure in the complex spectrum, which can be quantified from the averaged phase $\theta_\alpha$ of the $z_\alpha$. 
In particular, one considers the single-number indicator $\langle\cos\theta\rangle$.
A 2D Poisson distribution gives a flat profile of the $z_\alpha$, and a single-number indicator $\langle\cos\theta\rangle \simeq 0$.
A non-Hermitian RMT distribution in the Ginibre Ensemble of Gaussian non-Hermitian random matrices yelds a nontrivial distribution of the $z_\alpha$, and a single-number indicator $\langle\cos\theta\rangle \simeq 0.24$.
Throughout the manuscript, we use Eqs.~\eqref{eqs:ham_ratio} and \eqref{eqs:complex_ratio} to assess integrable or chaotic structure in real and complex spectra, respectively.
Moreover, for all the simulations presented in the main text and in the Supplementary Information, we used the \textit{QuantumToolbox.jl} package~\cite{mercurio_ARXIV_2025} available in Julia.

\section{II. Construction of the Liouvillian stripe}

We detail here the numerical procedure for the construction of Liouvillian stripes in chaotic open quantum systems.
We first diagonalize the Liouvillian superoperator obtaining its spectrum $\lambda_\alpha = \Gamma_\alpha + i\Omega_\alpha$ and right eigenoperators $\hat{\eta}_\alpha$,
\begin{equation}
    \mathcal{L}\hat{\eta}_\alpha = \lambda_\alpha\hat{\eta}_\alpha.
\end{equation}
We organize the Liouvillian spectrum in a collection of rectangular boxes of width $d$ and we gauge out the real part of the Liouvillian eigenvalues, so that $\lambda_\alpha\to\Omega_\alpha$.
We compute the Hamiltonian ratio $\langle r \rangle$ over the (re-ordered) set of real numbers $\Omega_\alpha$.
By sweeping over $d$, we identify the point $d_{\rm max}$ for which $\langle r \rangle$ is maximum.
The set of real eigenvalues $\Omega_\alpha$ and right eigenoperators $\hat{\eta}_\alpha$ identifies the Liouvillian stripe.
In all the chaotic systems that we analyze throughout the paper, $\langle r \rangle(d)$ always presents a clear maximum $\langle r \rangle(d_{\rm max})\simeq 0.51$ at a given $d_{\rm max}$.
This behavior can be qualitatively understood from the following argument. If $d$ is too large, we mix many different eigenvalues, effectively uncorrelated, and this yields a Poissonian statistics.
If $d$ is too small, the selected eigenvalues start to be distant on the vertical axis, and the spectral correlations are again destroyed.
In between these two extrema, there is an optimal point for which the real $\Omega_\alpha$ inherit the chaotic spectral structure of the complex $\lambda_\alpha$.
Moreover, we find that the width of the stripe always satisfies the condition, $d_{\rm max}\ll2\max(|\Omega|)$, making the stripe an effective one-dimensional object.
If the set of complex eigenvalues is Poissonian, the set of $\Omega_\alpha$ exhibits a 1D Poissonian statistics regardless of $d$. 

\begin{figure*}[t!]
\centering
\includegraphics[width=1\textwidth]{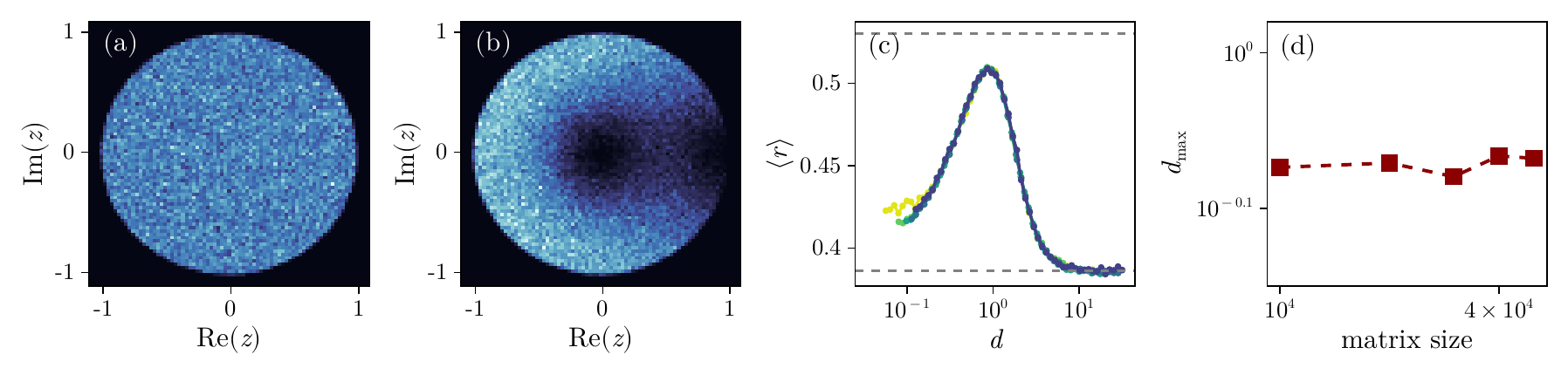}
\caption{Construction of the Liouvillian stripe in non-Hermitian random matrices. (a) and (b) Complex spacing ratios for $10^5$ uncorrelated complex random numbers in the interval $[0, 1]\times[0, 1]$ and for 3 random matrices with size $40000\times40000$, respectively.
(c) Hamiltonian ratio defined in Eq.~\eqref{eqs:ham_ratio} as a function of the stripe width $d$, for non-Hermitian random matrices with size $10000\times10000$, ..., $50000\times50000$ (from light green to dark blue).
Results have been obtained upon averaging over the disorder.
The gray-dashed lines indicates the values of $\langle r \rangle$ for uncorrelated random variables and for random matrices sampled from the Gaussian orthogonal ensemble.
(d) Scaling of the stripe width $d_{\rm max}$ for which $\langle r \rangle$ is maximum as a function of the size of the random matrices.}
\label{fig:random_matrices}
\end{figure*}

\begin{figure*}[t!]
\centering
\includegraphics[width=1\textwidth]{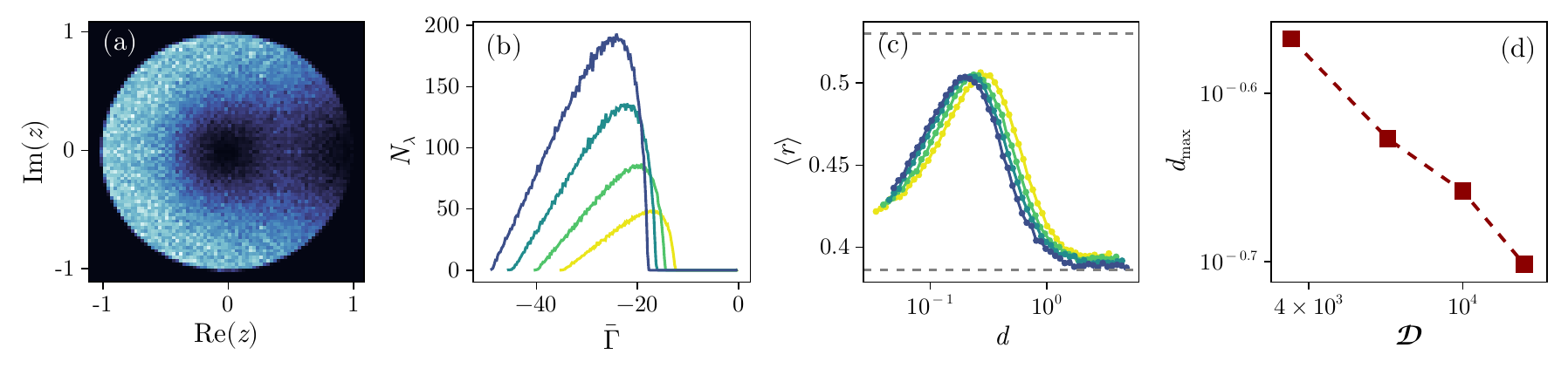}
\caption{Construction of the Liouvillian stripe in random Liouvillians. (a) Complex spacing ratios for $N_{\rm dis}=14$ random Liouvillians with Hilbert space size equal to $\mathcal{D}=120$. 
(b) Number of Liouvillian eigenvalues as a function of the decay rate $\textrm{Re}(\lambda)$ for $\mathcal{D}=60, 80, 100, 120$ (from light green to dark blue). Results have been obtained upon averaging over the disorder.
(c) Hamiltonian ratio defined in Eq.~\eqref{eqs:ham_ratio} as a function of the stripe width $d$ for the same system sizes and disorder averages of panel (b).
The gray-dashed lines indicates the values of $\langle r \rangle$ for uncorrelated random variables and for random matrices sampled from the Gaussian orthogonal ensemble.
(d) Scaling of the stripe width $d_{\rm max}$ for which $\langle r \rangle$ is maximum as a function of the dimension of the random Liouvillian.}
\label{fig:random_liouvillian}
\end{figure*}

\begin{figure*}[t!]
\centering
\includegraphics[width=1\textwidth]{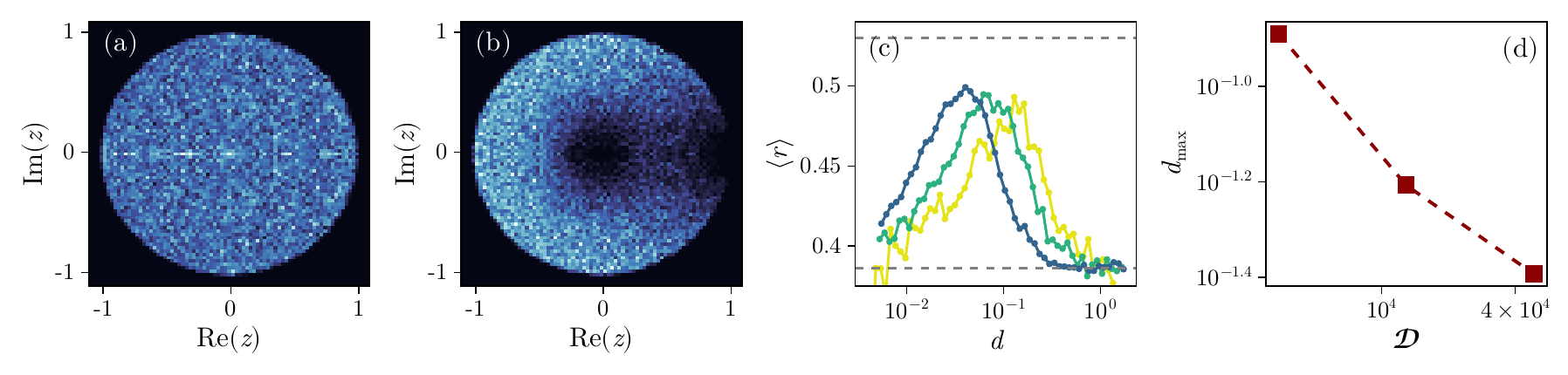}
\caption{Construction of the Liouvillian stripe for the driven-dissipative spin chain described by Eqs.~\eqref{eqs:spin_chain_1_ham} and \eqref{eqs:spin_chain_1_diss}. (a) and (b) Complex spacing ratios for a $N=9$ spin chain in the zero magnetization symmetry sector. In panel (a) we set $\Delta=h=0$, in panel (b) we set $\Delta=0.8$ and $h=1$. 
(c) Hamiltonian ratio defined in Eq.~\eqref{eqs:ham_ratio} as a function of the stripe width $d$ for $L=7, 8, 9$ spins (from light green to dark blue) for the quantum chaotic case.
The gray-dashed lines indicates the values of $\langle r \rangle$ for uncorrelated random variables and for random matrices sampled from the Gaussian orthogonal ensemble.
(d) Scaling of the stripe width $d_{\rm max}$ for which $\langle r \rangle$ is maximum as a function of the Liouvillian size, for the quantum chaotic case.}
\label{fig:spin_chain1}
\end{figure*}

\begin{figure*}[t!]
\centering
\includegraphics[width=1\textwidth]{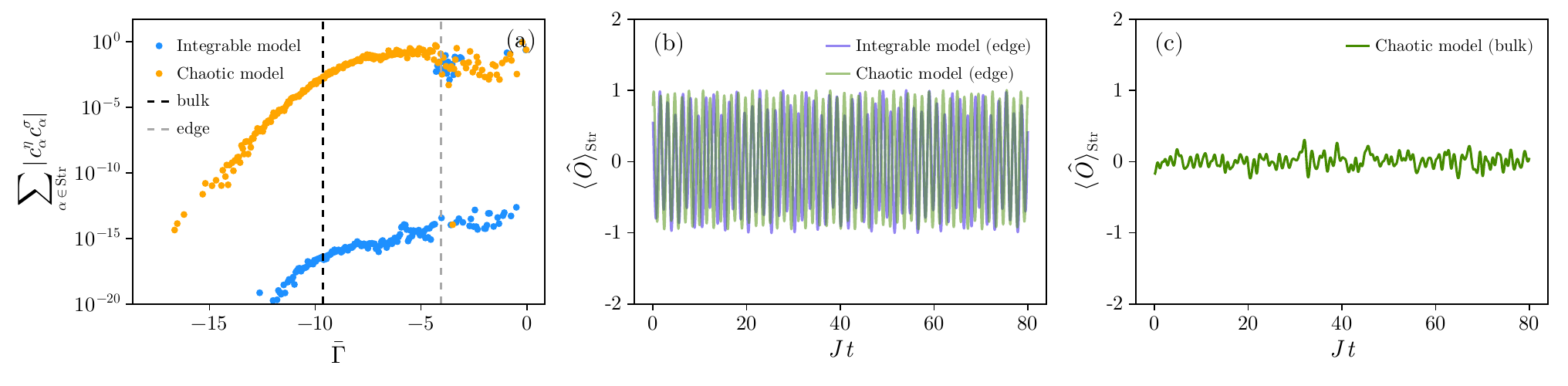}
\caption{Liouvillian ETH and dynamics for the local magnetization operator $\hat{\sigma}_3^z$. 
(a) Sum of the spectral coefficients over single stripes $\sum_{\alpha\in\textrm{Str}}|c_\alpha^\eta c_\alpha^\sigma|$, for the integrable (blue dots) and chaotic (orange dots). Each dot represents a single stripe and the width of the stripe is chosen according to the procedure described above.
(b) Renormalized dynamics of the local operator $\hat{O} = \hat{\sigma}^z_3$ over a stripe on the edge of the spectrum, according to Eq.~\eqref{eqs:dynaics_stripe} at a $\overline{\Gamma}$ indicated by the gray-dashed line in panel (a). 
The purple line refers to the integrable model, while the green line indicates the chaotic model.
(c) Renormalized dynamics over a stripe deep in the spectral bulk at a $\overline{\Gamma}$ indicated by the black-dashed line in panel (a).
Only the the dynamics for the chaotic model is reported here.
The values of $\Delta$ and $h$ are both 0 for the integrable model and $\Delta=0.8$ and $h=1$ for the chaotic model.
The other parameters are fixed as in Fig.~\ref{fig:ETH_spin1}.}
\label{fig:dynamics_app}
\end{figure*}

We test our hypothesis on several non-Hermitian system, proceeding by degrees of structure.
In Fig.~\ref{fig:random_matrices} we study the Liouvillian stripes in structureless non-Hermitian random matrices.
In Fig.~\ref{fig:random_matrices} (a) we plot the complex spacing ratios for $10^5$ independent complex random numbers uniformly sampled from the box $[0, 1]\times[0,1]$.
The Hamiltonian ratio $\langle r \rangle$ coincides with 0.386 regardless of $d$ (data not shown).
In Fig.~\ref{fig:random_matrices} (b) we plot the complex spacing ratios for 3 non-Hermitian Gaussian random matrices with size $40000\times40000$.
The Hamiltonian ratio for non-Hermitian random matrices with sizes $10000\times10000$, $20000\times20000$, ..., $50000\times50000$ as a function of the stripe width $d$ is reported in Fig.~\ref{fig:random_matrices} (c).
We observe how all the curves superimpose.
Finally, in Fig.~\ref{fig:random_matrices} (d) we plot the scaling of $d_{\rm max}$ with the size of the random matrices.
We observe that, for random matrices, the width of the Liouvillian stripe remains constant upon increasing the system size, whereas, constraining the spectrum in the box $[-0.5, 0.5]\times[-0.5, 0.5]$, it scales polynomially with the system size (data not shown).

Next, we construct the Liouvillian stripe for random Liouvillian systems, which we build up following the procedure detailed in the main text and in Refs.~\cite{denisov_universal_2019, sa_spectral_2020}.
The random Liouvillian can be viewed as a non-Hermitian random matrix with Lindblad structure, ensuring the existence of a steady state (whose properties have been studied in, e.g., \cite{sa_spectral_2020}) and complex conjugate eigenvalues (whose universal distribution has been studied in, \textit{e.g.}, \cite{denisov_universal_2019}), but it lacks any notion of locality and dimensionality.
In Fig.~\ref{fig:random_liouvillian} (a) we plot the complex spacing ratios of 14 random Liouvillians with Hilbert space dimension equal to $D = 120$. We choose the same parameters used in the main text: $r=6$, $\beta=2$, $g_{\rm eff} = 1.05$.
In Fig.~\ref{fig:random_liouvillian} (b) we plot the number of eigenvalues over the Liouvillian stripes, for random Liouvillians with Hilbert space dimensions $D = 60, 80, 100, 120$, as a function of $\overline{\Gamma}$, averaged over the same disorder realizations considered in the main text.
In Fig.~\ref{fig:random_liouvillian} (c) we show $\langle r \rangle$ as a function of $d$.
In this case, the position of the optimal width $d_{\rm max}$ [used for panel (b) also] moves to the left with $D$ and scales polynomially with the Liouville space dimension $\mathcal{D}=D^2$, as shown in In Fig.~\ref{fig:random_liouvillian} (d).

Finally, we construct the Liouvillian stripe for a structured driven-dissipative quantum spin chain. The Hamiltonian and jump operators are discussed in the main text.
As in the main text, we focus on the zero-magnetization symmetry sector.
In the absence of anisotropy ($\Delta=0$) and magnetic impurity ($h=0$), the model is integrable \cite{medvedyeva_exact_2016} and the complex spacing ratios distribution presented in Fig.~\ref{fig:spin_chain1} (a) for a $N=9$ chain displays the flat profile typical of uncorrelated random variables.
When the integrability-breaking terms proportional to $\Delta$ and $h$ are turned on, the Liouvillian becomes quantum chaotic, as showed by the distribution of the complex spacing ratios in Fig.~\ref{fig:spin_chain1} (b).
The Hamiltonian ratio for $N=7, 8, 9$ spins is displayed in Fig.~\ref{fig:spin_chain1} (c), and, similarly to the random Liouvillian, shifts to the left with the system size.
The scaling of $d_{\rm max}$ with the dimension of the Liouvillian $\mathcal{D}$ is displayed in Fig.~\ref{fig:spin_chain1} (d).

\section{III. Dynamics of other operators over the Liouvillian stripes}

In the main text, we studied the dynamics of the current operator $\hat{O} = i(\hat{\sigma}^+_1\hat{\sigma}^-_L - \hat{\sigma}^+_L\hat{\sigma}^-_1)$. 
Here we study the dynamics of the local magnetization in the bulk, $\hat{O} = \hat{\sigma}_3^z$ for the integrable ($\Delta=h=0$) and the chaotic ($\Delta=0.8$ and $h=1$) spin chain in Eqs.~\eqref{eqs:spin_chain_1_ham} and \eqref{eqs:spin_chain_1_diss}, respectively.
In Fig.~\ref{fig:dynamics_app} (a) we plot the sum of the spectral coefficients over the Liouvillian stripes, $\sum_{\alpha\in\textrm{\rm Str}}|c_\alpha^\eta c_\alpha^\sigma|$.
For the chaotic model (orange dots), the local operator has a support which spreads over the full spectrum, as suggested by the smooth and slow decay of the spectral coefficients.
For the integrable model (blue dots), the local operator has a support over only a few eigenvalues on the edge of the spectrum close to the steady state. 
In the bulk, the sum of the spectral coefficients suddenly drops to values below $10^{-15}$.
This implies that the full spectral bulk is not relevant for the dynamics of the given operator $\hat{O}$, and only eigenvalues close to the steady state and on the spectral edge determine the operator dynamics.
These considerations hold for many other local one- and two-site operators we considered.
Since only a few eigenstates participate in the dissipative time evolution of $\langle\hat{O}\rangle$, the model can not exhibit Liouvillian ETH, independently of its integrability.
Again, this is in analogy with the features of steady-state quantum chaos introduced in Ref~\cite{ferrari_dissipative_2025}.
In Fig.~\ref{fig:dynamics_app} (b) we plot the renormalized dynamics of $\hat{\sigma}_3^z$ over a stripe at the edge of the integrable and chaotic spectra, corresponding to a $\overline{\Gamma}$ indicated by the gray-dashed line in panel (a). 
In both the cases, only a few (less than 10) eigenvalues are involved in the stripe dynamics, and the renormalized dynamics given by Eq.~\eqref{eqs:dynaics_stripe} exhibits almost ideal oscillations bounded between 1 and $-1$.
In Fig.~\ref{fig:dynamics_app} (c) we plot the dynamics over a Liouvillian stripe deep in the chaotic spectral bulk, at a $\overline{\Gamma}$ indicated by the black-dashed line in panel (a).
Here, oscillations are markedly damped by the interference effect generated by all the chaotic eigenstates and eigenvalues in the Liouvillian stripe, similar to what we found in the main text for the current operator.

\end{document}